\documentclass[a4paper,12pt]{article}
\pdfoutput=1
\usepackage{jcappub}

\usepackage{graphics}

\usepackage{setspace}
\usepackage{setspace}
\usepackage{graphicx,color}
\usepackage{epstopdf}
\usepackage{setspace}
\usepackage{makeidx} 
\usepackage{lmodern}

\usepackage[T1]{fontenc}
\usepackage[utf8]{inputenc}
\usepackage{color}
\usepackage{booktabs}
\usepackage{units}
\usepackage{amsmath}
\usepackage{amssymb}
\usepackage{esint}
\usepackage{braket}

\usepackage{caption}
\usepackage{subcaption}
\usepackage[export]{adjustbox}
\pdfpageheight\paperheight
\pdfpagewidth\paperwidth

\def\bea{\begin{eqnarray}}  
\def\eea{\end{eqnarray}}

\def\be{\begin{equation}}
\def\ee{\end{equation}}

\def\H0{H_{0}}

\makeatletter

\newcommand{\Rmnum}[1]{\expandafter\@slowromancap\romannumeral #1@}
\makeatother

\usepackage{natbib}
\title{ Modified gravity, gravitational waves and the large-scale structure of the Universe: A brief report}

\author[a]{Ippocratis D.~Saltas} \author[b]{Luca Amendola} \author[c]{Martin Kunz} \author[d]{Ignacy Sawicki}

\affiliation[a, d]{CEICO, Institute of Physics of the Czech Academy of Sciences, Na Slovance 2, 182 21 Praha 8, Czechia}

\affiliation[b]{Institut für Theoretische Physik, Ruprecht-Karls-Universität Heidelberg,
Philosophenweg 16, 69120 Heidelberg, Germany}

\affiliation[c]{Départment de Physique Théorique and Center for Astroparticle Physics, Université de Genève, Quai E. Ansermet 24, CH-1211 Genève 4, Switzerland}

\keywords{dark energy, gravitational waves, large-scale structure of the Universe}

\abstract{The goal of this short report is to summarise some key results based on our previous works on model independent tests of gravity at large scales in the Universe, their connection with the properties of gravitational waves, and the implications of the recent measurement of the speed of tensors for the phenomenology of general families of gravity models for dark energy.}

\begin{document}

\maketitle
\flushbottom
\section{Introduction}
The physics behind the late-time acceleration of the Universe remains elusive. Although the $\Lambda$CDM model provides a good and simple phenomenological paradigm for its description, its fundamental understanding suffers. In particular, when attempting to explain the cosmological constant as the zero-point fluctuations of all fields in Nature, one faces a severe naturalness problem: the value of the observed (low-energy) cosmological vacuum is extremely sensitive to the details of the short-scale, UV physics \cite{Weinberg:1988cp, Bull:2015stt,Padilla:2015aaa}. In the absence of a UV completion of the standard model of particle physics, it is possible that the answer might hide in some new high energy physics yet to be discovered. 
\vspace{0.15cm}

A description of the accelerating Universe might also be embedded within an infrared modification of the standard gravitational paradigm of General Relativity (GR). This idea has sparked the construction of theories which modify the effective gravitational degrees of freedom at large scales, and are usually labelled as dark energy and/or modified gravity models (for detailed expositions, see e.g \cite{Clifton:2011jh, Amendola2010}). Typical examples are the general scalar-tensor (Horndeski and beyond as well as the so--called DHOST theories), vector-tensor (Einstein-Aether, generalised Proca) and massive/bi-metric theories of gravity. These are the kind of models that will concern us in the discussion to follow. 
\vspace{0.15cm}

The current major task of research in cosmology is the investigation of possible departures from the standard framework of GR. From an observational standpoint, the currently unknown detailed physics of the dark sector make this particularly challenging. This stems from the unobservability of the {\it individual} dark components under gravity probes, that is, the dust-like dark matter and the dark energy component. A way around to this hurdle is the use of parametrisations as proxies into the unknown physics. At the same time, it is highly desirable to have model-independent diagnostic tools at hand that could reveal unambiguous information, free from potential biases. Below we will explain how {\it the gravitational slip} parameter is a powerful model independent discriminator between large classes of gravity models in cosmology. We will further discuss its intimate relation to the properties of gravitational waves (GWs) \cite{TheLIGOScientific:2017qsa,Monitor:2017mdv}, that provides a phenomenological bridge between the physics of GWs and the large scale structure of the Universe. In light of the recent measurement of the speed of GWs, a reverse engineering allows to predict the consequences for large classes of gravity models at large scales. 
\vspace{0.15cm}

{\it The goal} of this short paper is to summarise some key results in the above directions based on our previous original works \cite{Amendola:2012ky, Motta:2013cwa, Saltas:2014dha, Sawicki:2016klv, Amendola:2017orw} \footnote{It is important to stress out here that, for a full list of references the reader should refer to the original works.} In Section \ref{sec:Obs} we briefly discuss the implications of the (un)observability of dark energy and dark matter for constraints on dark energy, and in Section \ref{sec:slip} we introduce the gravitational slip as a model independent observable quantity. In Sections \ref{MG-GWs} and \ref{sec:slip-GW} we discuss an important link between GWs and gravitational slip, while Section \ref{sec:Pheno} reports on the most recent phenomenological update on general gravity models after GW170817 and GRB 170817A.

\section{Preliminary notions} \label{sec:Defs}
The detailed structure and dynamics of different gravity models can differ drastically, however, their {\it phenomenological} effect on the physics of the Universe at the level of linear, scalar fluctuations may be simply summarised into {\it the following effects}: 
\vspace{0.2cm} 

{\bf i)} The change of the strength of gravity at large scales, parametrised by the {\it effective Newton's constant} $Y(t,k) \equiv -2k^2 M_{*}^{2} \Psi/(a^2 \rho_m \delta_m)$ \footnote{Here, we are using the notation and definitions of \cite{Bellini:2014fua}.}, where the matter density fraction is defined as $\delta_m \equiv \delta \rho_m/\rho_m$ and the effective Planck mass $M_{*}^{2}$ is an in principle time-dependent quantity. The immediate effect of $Y$ is to enhance ($Y > 1$) or weaken ($Y < 1$) the clustering strength of matter at large scales as compared to GR (where $Y = 1$). 
 \vspace{0.15cm}

{\bf ii)} A modification of the {\it weak lensing} effect, usually parametrised through the effective parameter $\Sigma(t,k) \equiv (1/2)(1+\eta)Y$, with $\eta \equiv \Phi/\Psi$.  In GR, as well as in dark energy models not affecting the dynamics of null geodesics, it is $\Sigma = 1$. 
\vspace{0.2cm}
 
Knowledge of these effective parameters for a gravity model defines its phenomenological footprint that can be tested with current and future galaxy/weak lensing surveys \cite{Amendola:2012ys}. In Section \ref{sec:Pheno} we discuss the updated phenomenology in this context for general models after the recent GW observations. 
\vspace{0.2cm} 

Now, a measurement of the above two effects allows for a reconstruction of a crucial quantity which will be central here: the {\it gravitational slip parameter} $\eta$ which is defined as the ratio of the two scalar, linear gravitational potentials $\eta(t,k) \equiv \Phi/\Psi$. As we will discuss later on, the generation of {\it gravitational slip}, that is, a difference in the magnitude of the two potentials ($\eta(t,k) \neq 1$), in the presence of perfect fluid matter will be the smoking gun of a modification of gravity at cosmological scales. In this sense, $\eta$ provides us with a key discriminator amongst large families of gravity models at large scales. Notice that, a scale-dependent $Y$ and/or $\eta$ implies its time-dependence, but the opposite is not necessarily true. Therefore, they constitute distinct effects from a phenomenological viewpoint.
\vspace{0.2cm} 

Large scale inhomogeneities are sourced by scalar fluctuations of the metric around the homogeneous and isotropic FLRW spacetime. For the perturbed line element we will be assuming that fluctuations are described according to $ds^2 = -(1 + 2\Psi({\bf x},t) )dt^2 + a(t)^2(1 - 2\Phi({\bf x},t))d{\bf x}^2$. Notice that, ``$k$" will usually stand for the wavenumber in Fourier space.

\section{What is the problem with the dark sector?} \label{sec:Obs} \footnote{This section closely follows the work of \cite{Amendola:2012ky, Motta:2013cwa}.}
The dominant force at large scales in the Universe is gravity. As such, our main observational window when extracting information about the dark sector is processes that involve gravitational interactions. However, our current knowledge about the properties of dark energy and dark matter, makes it difficult to infer information about one of the two dark components without making assumptions about the other. The root of the problem lies on what has been dubbed as the ``dark degeneracy" \cite{Kunz:2006ca}: Through the Einstein equations, geometry responds to the total energy-momentum content in the Universe, making it practically impossible to distinguish the particular properties of the individual dark components under pure gravity probes. 

Let us make things more concrete following the concept followed in our previous work \cite{Amendola:2012ky, Motta:2013cwa}. In a cosmological context, galaxies are the essential test particles: While their contribution to the total energy-momentum content of the Universe is negligible, their clustering and velocity field probe the physics of gravity at large scales. Both observables can be reconstructed in redshift and scale through the effect of {\it redshift space distortions}, without any assumptions on the underlying gravitational model according to 
\begin{equation}
\delta_{\text{gal}}^{z}(k,z,\mu)= \delta_{\text{gal}}(k,z)-\mu^{2}\frac{\theta_{\text{gal}}(k,z)}{H(z)},
\end{equation}
where $\mu$ is the direction cosine, and $\delta_{\text{gal}}^{z}(k,z,\mu)$ the galaxy density fraction in redshift space (see \cite{Amendola:2012ky} for more details). Therefore, the study of galaxy clustering in redshift space allows to map out the galactic density ($\delta_{\text{gal}}$) and velocity field respectively \cite{Kaiser}. 
The point where the dark degeneracy kicks in is when we attempt to use this information to infer the properties of the individual dark components. To understand why this is so, we remind that galaxies are usually seen as biased tracers of the underlying dark matter distribution, and for their density fraction one usually writes,
\begin{equation}
\delta_{\text{gal}} = b(z,k) \delta_{\text{DM}}. \label{eq:delta-gal}
\end{equation}
The bias function, $b(z,k)$, is unknown in a model-independent way, and so is the dark matter density field. The situation gets more complicated if we assume that dark energy has some clustering component contributing to the total matter density (there is no good reason to neglect this a priori!). In this case, $\Omega_{\text{DM}}\delta_{\text{DM}} \to \Omega_{\text{DM}}\delta_{\text{DM}} + \Omega_{\text{DE}}\delta_{\text{DE}}$, and the interpretation of $b(z,k)$ obviously changes too. At the level of the Poisson equation this translates to the fact that a non-standard effective Newton's constant would be unable to distinguish between clustering dark energy and/or modified gravity, since the potential responds to the product $G \cdot \delta_{\text{total}} \equiv Y \cdot \delta_{\text{DM}}$, with $G$ and $Y$ the bare and effective Newton's constants respectively. This is to be compared with a well known similar issue in astrophysics, when trying to extract the mass of a star. There, the directly measurable quantity is the product $G \cdot M$, and a measurement of the mass requires an assumption on $G$.

A similar problem emerges when studying the galactic velocity field ${\bf v}_{\text{gal}}$. Assuming that the weak equivalence principle holds, dark matter and baryons will fall under the same rate in a gravitational field (${\bf v}_{\text{gal}} = {\bf v}_{\text{DM}}$), then the energy-momentum conservation at the linear level tells us that,
\begin{equation}
\theta_{\text{gal}} = \theta_{\text{DM}} \simeq \delta_{\text{DM}}', 
\end{equation}
where $' \equiv d/dt$ and $\theta_{\text{gal}} \equiv {\bf \nabla} \cdot {\bf v}_{\text{gal}}$. This simple relation implies that measurements of the galactic velocity field can only allow for a reconstruction of the dark matter density if we input an initial condition at a sufficiently early time -- this is unknown in a model independent way. 

Below we will explain how the gravitational slip parameter $\eta  \equiv \Phi/\Psi$ is a powerful observable able to distinguish between large classes of models, that can be re-constructed in a model independent manner, and therefore, get around above obstacles.

\section{The gravitational slip and model independence} \label{sec:slip} \footnote{This section closely follows the original work presented in \cite{Amendola:2012ky, Motta:2013cwa}.}
The gravitational slip parameter $\eta$ is a direct measure of scalar (linear) anisotropic stress. At the linear level, the anisotropy constraint reads as
\begin{equation}
\Phi - \Psi = \sigma(t)  \Pi(t,k). \label{eq:AnisoEq-Gen}
\end{equation}
Here $\Pi$ is a functional of background quantities and linear perturbation variables (time and scale dependent), while $\sigma(t)$ depends on background quantities only \footnote{It is not hard to see that, for any model the anisotropy constraint can be always brought into this convenient form.}. Anisotropic stress therefore sources a gravitational slip, i.e $\eta \neq 1$. At late times, the free streaming of neutrinos makes a negligible contribution to the anisotropic pressure part of the total energy-momentum in the Universe. Instead, its only source in the presence of perfect fluid matter is a possible modification of gravity. For minimally coupled scalar field theories (quintessence, k--essence) or $\Lambda CDM$, it is $\eta = 1$, while $\eta \ \neq 1$ in non-minimally coupled models such as $f(R)$/Brans--Dicke. Obviously this discussion needs to be made on firm grounds -- the distinction between different models will be made precise in section \ref{MG-GWs} when we will introduce our working definition of modified gravity. 

We will now show how $\eta$ can be reconstructed in a model-independent manner, closely following the original work of \cite{Amendola:2012ky, Motta:2013cwa}. As explained earlier, in the standard context, galaxies in cosmology are treated as biased tracers of the underlying dark matter distribution. However, the view we will adopt here is different -- instead, we will treat galaxies as tracers of the local gravitational potentials without any reference to the underlying distribution of the dark components. 

Let us start by assuming that galaxies move on geodesics, a valid assumption as long as non-linearities associated with e.g spacetime effects or galaxy-galaxy interactions can be neglected. In addition, it uniquely fixes the conformal frame, an issue inherit in theories with different conformal representations (the so--called Jordan/Einstein frame). The resulting geodesic equation reads as
\begin{equation}
\left(a^{2}\theta_{\text{gal}}\right)'=a^{2}Hk^{2}\Psi\,.\label{eq:rsd}
\end{equation}
Therefore, the observable galactic velocity field $\theta_{\text{gal}}$ can be simply seen as a measurement of the gravitational potential $\Psi$ at different redshifts and scales. 

Weak lensing is a relativistic effect describing the lensing of light as it passes close to large matter inhomogeneities such as a galaxy cluster, traditionally used as a probe of the underlying dark matter distribution. However, we can instead use it as a direct measurement of the lensing potential only according to the Poisson-like equation
\begin{equation}
k^{2}\Phi_{\text{lens}} \equiv k^{2}(\Psi + \Phi) \equiv -\frac{3(1+z)^{3}}{2E^{2}}L\, , \label{eq:Phi-Lens}
\end{equation}
where $L \equiv \Omega_{\text{m}0}GY(1+\eta)\delta_{\text{m,0}}$. Therefore, tomographic weak lensing observations provide a measurement of the linear combination of $\Phi$ and $\Psi$. It is clear that, had we tried to map the lensing potential to the parameters of the dark sector we would had run into the usual problem of dealing with in principle unknown quantities ($\Omega_{\text{m}0}$, $\delta_{\text{m,0}}$). 

Now, combining equations (\ref{eq:rsd}) and (\ref{eq:Phi-Lens}) we can derive a relation for the gravitational slip solely in terms of observable quantities as
\begin{equation}
\eta= \Phi/\Psi=\frac{3(1+z)^{3}}{2E^{2}\left(\nicefrac{R'}{R}+2+\frac{E'}{E}\right)} \frac{L}{R}-1\, ,\label{eq:eta_obs}
\end{equation}
where we introduced the velocity-field observable $R \equiv - \theta_{\text{gal}}/H(z)$, and the background evolution $E(z) \equiv H(z)/H_0$, while $L$ was defined after (\ref{eq:Phi-Lens}). Relation (\ref{eq:eta_obs}) provides a model-independent expression for the gravitational slip parameter first derived in \cite{Amendola:2012ky}. Given a background expansion history (through $E(z)$), it allows for the reconstruction of $\eta$ in redshift and scale from galaxy kinematics and weak lensing observations. For a recent reconstruction of $\eta$ from observations according to (\ref{eq:eta_obs}), see \cite{Pinho:2018unz}.

\section{Modified gravity and gravitational waves} \label{MG-GWs}
The distinction between dark energy and modified gravity is sometimes blurry. For a meaningful and unambiguous discussion on how to discriminate between $\Lambda$CDM/dark energy and modified gravity models, we need to have a consistent definition of what we mean by dark energy and modified gravity in a given context. Here, following \cite{Saltas:2014dha} we will define a modification of gravity as any theory that changes the propagation of the dynamical degrees of freedom of GR, i.e the gravitational waves (GWs). If we assume that GWs are far away from their source (plane-wave approximation), we can write for their propagation
\begin{equation}
h''_{ij}+ \big(2+\nu(t) \big)H h'_{ij} + c_\text{T}^2(t) k^2h_{ij}+a^2\mu^{2}h_{ij}=a^2\Gamma(t)\gamma_{ij}(t)\,,\label{eq:GWeq}
\end{equation}
where the field $h_{ij}$ is the spin-two fluctuation of the metric. FLRW symmetry ensures that, general extensions of GR can modify the propagation of GWs only through one of the following parts: the friction ($\nu(t)$), the propagation speed ($c_\text{T}^2$), mass ($\mu^2$) and source term ($\Gamma\gamma_{ij}$). Let us now introduce the models that we will be considering, which we conveniently divide into models introducing a new dynamical field of spin-0 (scalar-tensor), spin-1 (vector-tensor) and a spin-2 (massive/bi-gravity) respectively:
\\

{\it Scalar-tensor theories:} These start with Brans-Dicke gravity and its generalisations, i.e Horndeski \cite{Horndeski:1974wa}, Beyond Horndeski and DHOST theories \cite{horndeski, BenAchour:2016fzp, Langlois:2017mxy}. In Brans--Dicke gravity (or its sibling $f(R)$ theory), the scalar couples to curvature via a conformal coupling. This is radically generalised in Horndeski gravity and its extensions, where the scalar is kinetically coupled to curvature in a non-trivial manner. The Horndeski theory in particular, is the most general second-order theory that can be constructed out of a scalar and the metric, while its higher-order extensions remove the Ostrogradsky ghost through appropriate constraints. In general, the presence of a non-minimal coupling between curvature and the scalar field will induce a running of the effective Planck mass which in turn modifies the friction term in (\ref{eq:GWeq}) with $\nu \equiv H^{-1}d \ln M(t)^2/dt$. At the same time, the existence of a non-trivial kinetic mixing between the scalar and the metric alters the light-cone structure of GWs with respect to that of photons, leading to $c_{T}^2 \neq 1$. All other modifications in (\ref{eq:GWeq}) equal zero for this class of models. 
\vspace{0.2cm}

{\it Vector-tensor theories:} This class of theories includes Einstein-Aether (EA) \cite{Jacobson:2000xp}, which is related to the low-energy limit of Horava-Lifsitz gravity \cite{Blas:2010hb}. The extension of the old (massive) Proca theory is described by the generalised Proca theories \cite{Tasinato:2014eka,Heisenberg:2014rta}, corresponding to the general covariant theory of a massive vector on a curved background. All of these vector-tensor theories only modify the speed of GWs in (\ref{eq:GWeq}), $c_{T}^2 \neq 1$, related to the non-trivial derivative interactions between the dynamical vector field and the metric. 
\vspace{0.2cm}

{\it (Massive)Bi-gravity theories:} This family of theories describes the dynamics of two, non-minimally interacting coupled spin-two fields \cite{Hassan:2011zd}. The non-minimal character of the interactions inevitably gives a mass to the standard (GR's) graviton, $\mu^2 \neq 0$. What is more, the second graviton sources the evolution of the standard one through the term on the r.h.s of (\ref{eq:GWeq}), while all other terms are standard. 
\vspace{0.1cm}

Let us note here that, our later discussion does not necessarily assume that the new degree of freedom is responsible for accelerating the Universe -- in this sense, one can test them as effective extensions of GR, beyond the problem of dark energy.

\section{The link between gravitational slip and gravitational waves} \label{sec:slip-GW}
\subsection{Non-standard propagation of gravitational waves implies a gravitational slip}
Let us assume that gravity is modified, according to its definition in Section \ref{MG-GWs}. This means that the propagation of GWs will be different to GR in one way or the other according to (\ref{eq:GWeq}).
In \cite{Saltas:2014dha} it was shown that a direct link exists between the propagation of GWs at any scale in the Universe and the existence of gravitational slip (or equivalently, anisotropic stress) in the presence of perfect fluid matter at large scales:
\vspace{0.2cm} 

{\it For general families of theories beyond GR introducing an extra spin-0, spin-1 and spin-2 two field respectively, the theory-space parameters that enter the coefficient of anisotropic stress, are exactly those that modify the propagation of GWs at any scale.}
\vspace{0.2cm} 

In view of the anisotropy and the GW propagation equation, (\ref{eq:AnisoEq-Gen}) and (\ref{eq:GWeq}) respectively, this translates to the theory-space parameters entering $\sigma$ to be the same modifying the propagation of GWs. Mathematically, this link intimately relates to the spatial traceless nature of both types of fluctuations. For the sake of an example, let us consider the Einstein-Aether vector-tensor theory. The free theory-space parameters are the three constants, $\beta_1, \beta_2, \beta_3$, associated with three operators, constructed out of products of the gradients of the vector field $\nabla_\mu u_\nu$ and the spacetime metric (see \cite{Jacobson:2000xp}). Within a standard kinematic decomposition of the vector's gradient into an acceleration, shear, expansion and twist terms respectively, the coefficient of the shear is $c_\sigma = \beta_1 + \beta_3$. It then turns out that in view of the anisotropy constraint, it is $\sigma = -c_\sigma$, while GWs acquire a non-standard propagation speed $c_{T}^2 = \left( 1 + c_{\sigma} \right)^{-1}$. It is straightforward to see the similar connection within the rest of the scalar-tensor or bi-gravity models described earlier. 
\vspace{0.2cm} 

{\it What is the significance of this link ?} The short answer is that it brings {\it complementary} information about gravity. Put differently, it establishes a phenomenological bridge between two apparent disconnected regimes: It connects the physics of large-scale structures in the Universe with the propagation of GWs at any scale. In this sense, it connects the phenomenology of spin-0 with spin-2 fluctuations in the Universe and allows to translate constraints from GW observations to constraints of gravity at cosmological scales. This will be discussed in Section \ref{sec:Pheno} based on the recent measurement of the speed of GWs. 

\subsection{Does the existence of gravitational slip imply modified gravity?} \label{sec:No-slip}
In the previous discussion we had $\eta \neq 1$ in the presence of perfect-fluid matter, assuming that gravity is modified. It is equally important to ask the opposite question: If there is no gravitational slip in the presence of perfect fluid matter (i.e $\eta = 1$), does this necessarily imply that gravity is not modified at large scales? In other words, {\it could a modified gravity theory configure itself in such a way that it hides itself at the level of linear large-scale structures?} Notice that, we are interested in a dynamical ``shielding", that is, configurations which preserve the no-slip along the time evolution of the model.

This question was explored in \cite{Sawicki:2016klv} for Horndeski, Einstein-Aether and bi-metric gravity, and the short answer is that, {\it it would be either impossible or highly unnatural.} The idea behind the analysis pursued is to show whether an initial configuration with no slip ($\eta = 1$) at some given time and {\it all} scales, can be preserved under the dynamics of the model. It turns out that, for the general classes of models as described in Section \ref{MG-GWs}, in the case where this might be in principle possible to happen (e.g in Horndeski gravity) this corresponds to strongly-coupled or similar pathological configurations \footnote{A sketch of the original analysis pursued in \cite{Sawicki:2016klv} is as follows. Starting from the anisotropy constraint, $C[{\bf X}] \equiv \Phi - \Psi = {\bf A} \cdot {\bf X}$, one requires that at some particular time and at all scales the anisotropic stress vanishes (such a configuration can be always found), i.e $C[{\bf X}] = 0$ (``no-slip condition"). The vector ${\bf X}$ represents the four phase space variables (second-order dynamics), while the vector ${\bf A}$ background-dependent coefficients. The existence of dynamical shielding then depends on whether the time evolution of the theory preserves the original no-slip condition along the equations of motion (``$\approx$") i.e whether $d^{n} C/dt^2 =  (d^{n} {\bf A}/dt^2)  \cdot {\bf X}  \approx 0$, with $n = 1,2,3$ since the phase space is four dimensional.}.
 From an observational point of view, this result suggests that a possible no-detection of gravitational slip (i.e $\eta = 1$), would allow us to be sufficiently confident that gravity is not modified. It should be stressed that by the moment one relaxes the requirement that $\eta = 1$ at all scales, it is possible to find viable configurations that yield $c_{T}^2 = 1$ and $\eta = 1$ in the quasi-static regime. This scenario has been introduced in \cite{Linder:2018jil} and has been dubbed as No-slip gravity.

\section{Dark energy phenomenology after GW170817} \label{sec:Pheno} \footnote{This section closely follows the results presented in \cite{Amendola:2017orw}.}
Recently, the speed of GWs was measured from an observation of a merging neutron star and an electromagnetic counterpart, yielding the extremely strong constraint $| c_T/c - 1| \leq 10^{-15}$ \cite{TheLIGOScientific:2017qsa}. The result provided a confirmation of GR's prediction, and placed the strongest constraints on the theory space of generic modified gravity models so far. The observation has essentially ruled out all gravitational interactions which change the light cone structure of GWs -- these are typically terms which mix kinetically the scalar with the metric, beyond the simple conformal coupling (Brans--Dicke/$f(R)$). In light of the previously discussed link between GWs and the large-scale structure of the Universe, this new constraint can be translated to phenomenological predictions at large scales. 
Below we discuss the implications for the theory space of different families of models following the results and notation of the original work \cite{Amendola:2017orw}. 
\\

{\it Scalar-tensor theories:} For Horndeski theories, the surviving theory space is described by 
\begin{equation}
\mathcal{L} = \frac{f(\phi)}{2} R + K(X,\phi) - G(X,\phi) \Box \phi. \label{Horndeski-surv}
\end{equation}
Notably, the only non-minimal coupling between scalar and curvature allowed is of the conformal type. The second term in (\ref{Horndeski-surv}) encompasses the k--essence/quintessence theory, while the third one the so--called kinetic gravity braiding  \cite{Deffayet:2010qz}. At the linear level of scalar perturbations, the theory (\ref{Horndeski-surv}) is described by three functions of time, namely the so--called kineticity ($\alpha_K(t)$), braiding ($\alpha_{B}(t)$) and Planck mass rate ($\alpha_{M}(t)$). The first one measures the amount of fluctuations of the scalar field, the second the braiding between scalar and metric fluctuations, while the third the rate of change of the evolving Planck mass. In the quasi-static limit, it turns out that the effective Newton's constant for the theory (\ref{Horndeski-surv}) reads as
\begin{equation}
	Y = 1 + \frac {(\kappa+\alpha_M)^2}{2N}\,,\quad
\end{equation}
with $Y$ defined in Section \ref{sec:Defs}. Here, $\kappa \equiv \alpha_M + \alpha_B$ is sourced only through the third operator in (\ref{Horndeski-surv}) and is zero otherwise, since $\alpha_M = -\alpha_B$ for Brans-Dicke/$f(R)$ and $\alpha_M = 0 = \alpha_B$ in minimally-coupled models. $N$ is associated with the sound speed of the scalar and is required to be positive. For the gravitational slip and lensing parameter one finds that,
\begin{equation}
\eta -1 = -\frac{2\alpha_M(\kappa+\alpha_M)}{2N+(\kappa + \alpha_M)^2}, \; \;  \, \; \; \Sigma \equiv  \frac{Y(\eta + 1)}{2}.
\end{equation}
\\
From these relations it is straightforward to conclude for the {\it updated phenomenology} of scalar-tensor theories at large scales: The strength of clustering can be only made stronger than or equal to that in GR, that is, $Y \geq 1$. The lensing is the same as in GR ($\Sigma = 1$) as soon as the third operator in (\ref{Horndeski-surv}) (braiding term) is absent ($\alpha_B = -\alpha_M$ or $\alpha_B = 0 = \alpha_M$) -- this is to be expected, as conformally- or minimally-coupled models do not affect null geodesics. However, as soon as the braiding term is included it is $\Sigma \neq 1$. As regards the slip parameter $\eta$, that can be on either side of its GR value, i.e $\eta < 1$ or $\eta \geq 1$.

Let us close with a short comment on the higher-order extension of Horndeski theories, the so--called Beyond Horndeski model. As was shown in \cite{Creminelli:2017sry,Sakstein:2017xjx}, there exists a combination on the free theory functions that satisfies the GW speed constraint. At the linear level of perturbations the theory introduces a new free function ($\alpha_H$), and the understanding of its updated phenomenological implications require a separate study left for future work. 
\\

{\it Vector theories:}
Let us start with Einstein-Aether theories. The theory introduces three constant parameters $\beta_i$ associated with each operator in the action. The requirement of $c_{T}^2 = 1$ imposes that $\beta_1 = -\beta_3$, and at the same time eliminates the vector's anisotropic stress contribution, fixing $\eta = 1$. For the effective Newton's coupling, it is $Y = (1+3\beta_2)/(1-\beta_1)$. Given that stability requires $\beta_2, \beta_1 > 0$, then $Y \geq 1$, as was the case in Horndeski models. 

On the other hand, generalised Proca theories are characterised by six free functions at the level of the action, and similar to Horndeski theory, these combine to form a set of time-dependent functions that control scalar linear perturbations. The terms in the action that affect the GW propagation are those that kinetically mix the vector with the metric -- It turns out that, for the associated free functions of these operators, $G_4=\text{const}$ and $G_5=0$ \cite{Baker:2017hug}. For this case the gravitational slip parameter becomes always equal to unity $\eta = 1$,
while for the effective Newton's coupling (within the quasi--static approximation) we find that,
\begin{equation}
	Y =1+\frac{-w_3w_2^2}{N}  \geq 1,
 \end{equation}
where we used the notation of \cite{DeFelice:2016uil} for the background functions $w_{i}(t)$. We further defined the quantity $N \equiv 2\mu_2/\phi +w_3w_2^2$, which is related to the sound speed of the new degree of freedom, and the functions $w_2, w_3, \mu_2$ are defined in \cite{DeFelice:2016uil}. Investigation of the explicit expression for $w_3$, along with the requirement of absence of ghost instabilities implies that $w_3 < 0$ and on similar grounds, the stability of scalar fluctuations requires that $N > 0$. From these, it follows for this case too that, $Y \geq 1$, that is, the clustering of matter can be made either equal or stronger compared to GR for these models ($Y \geq 1$), similar to the previous models. 

As we discussed in Section \ref{MG-GWs}, {\it massive/bi-gravity theories} predict that GWs propagate with $c_\text{T}^2 = 1$, hence the recent measurement has no implications for their theory space. 
\vspace{0.25cm} 

{\it Consistency conditions:} Based on the above, and working within the small-scale limit of the quasi--static approximation (i.e $k \to \infty$), we can extract two new consistency conditions for {\it scalar--tensor theories} at large scales \footnote{For minimally-coupled scalar-tensor models it is easy to see that $\eta  = 1$ and $Y > 1$. Notice that here, we do not consider the possibility of special configurations such as the scenario of No-Slip gravity of \cite{Linder:2018jil} mentioned in Section \ref{sec:No-slip}.}:
\vspace{0.2cm}

1. $\eta < 1$ and $\Sigma = 1$: An imprint of a conformally coupled scalar field (Branks-Dicke/$f(R)$).  \\

2. $\eta > 1$ and/or $\Sigma \neq 1$: An imprint of the presence of a kinetic braiding term in the action, in addition to the conformal coupling to curvature. \\
\\
For {\it vector models}, an important corollary is the characteristic combination $\eta = 1$ and $Y \geq 1$. 

It is interesting to remark that future direct observations of GWs can probe the running rate of the Planck mass in the surviving scalar-tensor sector as was argued in \cite{Belgacem:2017ihm, Amendola:2017ovw}. 

\begin{table}[ht] 
\centering
\caption{Phenomenological implications for the gravitational slip and lensing parameters $\eta$ and $\Sigma$ respectively, after the recent measurement of the speed of GWs (see Section \ref{sec:Pheno}) \cite{Amendola:2017orw}. The first two rows describe the surviving theory space of Horndeski models, while the second and third one the Einstein--Aether and generalised Proca vector theories respectively. In all models, it now turns out that for the effective Newton's coupling at large scales it is $Y \geq 1$, i.e clustering of matter is equal to or stronger than in GR. }\label{table-pheno}
\begin{tabular}[t]{lccccc}
\hline
{\bf Model}  &$ \eta \equiv \Phi/\Psi $& $\Sigma \equiv (1/2) (\eta + 1)Y$ &  $Y $  \\
\hline \\
\vspace{0.2cm}
$ M_{p}^2 R + K(\phi,X)$ & $1$ & $1$ & $Y \geq 1$\\
\vspace{0.2cm}
$f(\phi) R + K(\phi,X)$ & $\leq 1$ & $1$ & $Y \geq 1$\\
\vspace{0.2cm}
$f(\phi) R + K(\phi,X) + G(X,\phi) \Box \phi$   &$ <1$ or $\geq 1$& $\neq 1$ & $Y \geq 1$ \\
\vspace{0.2cm}
Einstein-Aether &  $ = 1$  & $\neq 1$ & $Y \geq 1$ \\
\vspace{0.2cm}
Generalised Proca &$ = 1$& $\neq 1$  & $Y \geq 1$ \\
\hline
\end{tabular}
\end{table}%
\section{Summary}
Testing the standard gravity paradigm of General Relativity is a primary goal of cosmology and astrophysics. The recent detection of gravitational waves (GWs) provided an exciting new observational window in this direction, complementary to current and future cosmological observations. In this short paper we summarised key results from our previous works on the significance and the construction of model independent tests of gravity at cosmological scales, their intimate relation with the properties of GWs and the most recent phenomenological implications of GW observations for dark energy models. 

In particular, we focused on the gravitational slip parameter as a key discriminator of large classes of gravity models that can be reconstructed in a model independent manner from redshift space distortions and weak lensing observations. The recent measurement of the speed of GWs placed surprisingly stringent constraints on the theory space of gravity models, and allowed for robust cosmological predictions that can be accurately tested with the future Euclid satellite mission \cite{Amendola:2012ys}. The new predictions and consistency conditions for the gravitational slip, weak lensing and effective gravity strength parameter summarised in Section \ref{sec:Pheno}, provide a challenge for the surviving models that will be tested with future surveys. Finally, future GW observations will further allow to test the surviving scalar-tensor sector through the induced running of the effective Planck mass in these theories. 

\acknowledgments
I.D.S and I.S are supported by ESIF and MEYS (Project CoGraDS --  \\ CZ.02.1.01/0.0/0.0/15\_003/0000437). This report is based on a talk given by I.D.S at the 15th Marcel Grossmann meeting -- I.D.S would like to thank Luca Amendola and Joan Sol\'a for the invitation. 

\bibliographystyle{utcaps}
\bibliography{AnisoRefs,WD-Bib}

\end{document}